\documentclass[sigplan,preprint,nonacm]{acmart}
\usepackage{booktabs}
\usepackage{graphicx}
\usepackage{xspace}
\usepackage[ruled,linesnumbered,vlined]{algorithm2e}
\SetAlFnt{\footnotesize}\SetAlCapFnt{\footnotesize}\SetAlCapNameFnt{\footnotesize}
\usepackage{microtype}
\usepackage{tcolorbox}
\newtcolorbox{findingbox}{
  colback=black!4, colframe=black!40, boxrule=0.5pt, arc=1.5pt,
  left=4pt, right=4pt, top=2pt, bottom=2pt, boxsep=0pt,
  before skip=3pt plus 1pt, after skip=3pt plus 1pt}
\newcommand{\finding}[2]{%
\begin{findingbox}\small\textbf{Finding (#1):} #2\end{findingbox}}
\newcommand{\AbCached}{68.9}
\newcommand{\AbNocache}{13.1}
\newcommand{\AbSpeedup}{5.3}
\newcommand{\AccPerTok}{1585}
\newcommand{\AcceptedB}{299}
\newcommand{\BwAdopt}{327}
\newcommand{\BwHbm}{2938}
\newcommand{\BwHostAlloc}{330}
\newcommand{\BwLink}{368.5}
\newcommand{\BwMalloc}{23}
\newcommand{\BwRunDev}{0.8}
\newcommand{\BwStream}{346}
\newcommand{\ColCLru}{71}

\newcommand{\ColHLru}{70}
\newcommand{\ColHPin}{45}

\newcommand{\ColMPin}{21}
\newcommand{\ColPLru}{74}
\newcommand{\ColPPin}{57}
\newcommand{\DAlphaHi}{0.93}
\newcommand{\DAlphaLo}{0.86}

\newcommand{\DJacHi}{0.085}
\newcommand{\DJacLo}{0.032}
\newcommand{\DLruB}{48}
\newcommand{\DStaticB}{35}
\newcommand{\DWOneHi}{31}
\newcommand{\DWOneLo}{26}

\newcommand{\DWSixtyHi}{85}
\newcommand{\DWSixtyLo}{83}
\newcommand{\DepthGradient}{4.6}
\newcommand{\DepthTopFirst}{3.9}
\newcommand{\DepthTopLast}{10.7}
\newcommand{\SingleTokA}{0.94}
\newcommand{\SingleTokC}{1.31}
\newcommand{\DepthTopMax}{18.0}
\newcommand{\DepthUniqMax}{890}
\newcommand{\DepthUniqMin}{646}
\newcommand{\DiskGBps}{27.0}
\newcommand{\EfivePerTokA}{0.82}
\newcommand{\EfivePerTokC}{1.30}
\newcommand{\EfiveVsA}{-13}
\newcommand{\EfiveVsC}{0}
\newcommand{\EngSpread}{2}
\newcommand{\EsixDC}{282}
\newcommand{\EsixDH}{296}
\newcommand{\EsixDP}{294}
\newcommand{\EsixHi}{1.10}
\newcommand{\EsixKC}{256}
\newcommand{\EsixKH}{273}
\newcommand{\EsixKP}{265}
\newcommand{\EsixLo}{1.09}
\newcommand{\EsixPairsC}{3}
\newcommand{\EsixPairsH}{3}
\newcommand{\EsixPairsP}{3}
\newcommand{\EsixTotHi}{1.04}
\newcommand{\EsixTotLo}{1.03}
\newcommand{\EsixXC}{1.10}
\newcommand{\EsixXH}{1.09}
\newcommand{\EsixXHiC}{1.10}
\newcommand{\EsixXHiH}{1.10}
\newcommand{\EsixXHiP}{1.11}
\newcommand{\EsixXLoC}{1.10}
\newcommand{\EsixXLoH}{1.06}
\newcommand{\EsixXLoP}{1.10}
\newcommand{\EsixXP}{1.10}
\newcommand{\EtwoBRand}{1.80}
\newcommand{\EtwoHyC}{1.11}

\newcommand{\EtwoPcC}{1.13}

\newcommand{\EtwoPcDiskC}{7.2}

\newcommand{\EtwoPcSvcC}{75.3}
\newcommand{\EtwoPcSvcD}{55.4}
\newcommand{\EtwoPinC}{1.04}

\newcommand{\EtwoPinDiskC}{7.4}

\newcommand{\EtwoPinHitC}{73.2}

\newcommand{\EtwoPinPreC}{256}

\newcommand{\EtwoPinSvcC}{74.6}
\newcommand{\EtwoPinSvcD}{57.0}
\newcommand{\EtwoTaxC}{1.09}
\newcommand{\EtwoTaxD}{1.11}
\newcommand{\EtwoVmLckC}{238}

\newcommand{\ExpertMB}{17.55}
\newcommand{\ExpertsPerStep}{1663}

\newcommand{\GsGlobalGapMax}{2.7}
\newcommand{\ItersB}{285}
\newcommand{\KAlphaHi}{1.04}
\newcommand{\KAlphaLo}{1.00}
\newcommand{\KBAccepted}{1078}

\newcommand{\KBCbTwoFiveFour}{64}
\newcommand{\KBCbTwoFiveOne}{78}
\newcommand{\KBDedup}{1.08}
\newcommand{\KBFixedB}{125}
\newcommand{\KBGBStep}{108}
\newcommand{\KBPerTokRatio}{1.45}

\newcommand{\KBeladyB}{62}
\newcommand{\KChatLruSixteen}{19.4}
\newcommand{\KChatStaticSixteen}{15.5}
\newcommand{\KCovHi}{69}
\newcommand{\KCovLo}{64}
\newcommand{\KJacHi}{0.188}
\newcommand{\KJacLo}{0.101}
\newcommand{\KLfuEight}{12.7}
\newcommand{\KLruB}{44}
\newcommand{\KLruEight}{0.5}
\newcommand{\KStar}{1.06}
\newcommand{\KStaticB}{23}
\newcommand{\KStaticEight}{9.8}
\newcommand{\KTopEightHi}{16}
\newcommand{\KTopEightLo}{13}
\newcommand{\KWEightHi}{59}
\newcommand{\KWEightLo}{57}
\newcommand{\KWOneHi}{36}
\newcommand{\KWOneLo}{32}

\newcommand{\KWSixtyHi}{82}
\newcommand{\KWSixtyLo}{80}
\newcommand{\LongAlphaHi}{1.71}
\newcommand{\LongAlphaLo}{1.54}
\newcommand{\LongCovHi}{81}
\newcommand{\LongCovLo}{73}

\newcommand{\MFullMs}{20.6}
\newcommand{\MPoolGB}{17.6}

\newcommand{\MSqueezeMs}{26.3}
\newcommand{\McAmpA}{0.99}
\newcommand{\McAmpC}{1.22}
\newcommand{\McAmpD}{1.50}
\newcommand{\McAmpE}{2.22}

\newcommand{\McInvErrA}{+0.0}
\newcommand{\McInvErrC}{+5.0}
\newcommand{\McInvErrD}{-6.7}
\newcommand{\McInvErrE}{-9.7}
\newcommand{\McNomA}{5.3}
\newcommand{\McNomC}{8.1}
\newcommand{\McNomD}{11.6}
\newcommand{\McNomE}{15.1}
\newcommand{\MgAmp}{2.6}

\newcommand{\MgOffDisk}{14}
\newcommand{\MgOffDiskHi}{19.0}
\newcommand{\MgOffDiskLo}{13.8}

\newcommand{\MgOffSpread}{38}
\newcommand{\MgOnDisk}{36}

\newcommand{\MgOnScanHi}{6.9}
\newcommand{\MgOnScanLo}{5.5}
\newcommand{\PfAdv}{1.077}
\newcommand{\PfAdvP}{1.131}
\newcommand{\PfBlk}{1.151}
\newcommand{\PfBlkP}{1.153}
\newcommand{\PfGain}{5.0}
\newcommand{\PfGainP}{0.3}
\newcommand{\PfNone}{1.134}

\newcommand{\PfSubm}{5{,}278}
\newcommand{\PilotLayerHi}{84}
\newcommand{\PilotLayerLo}{1}

\newcommand{\PilotRecall}{64.7}
\newcommand{\PoolTB}{1.45}
\newcommand{\ProbeCold}{27.0}
\newcommand{\ProbeWarm}{61.4}
\newcommand{\QAlphaHi}{1.26}
\newcommand{\QAlphaLo}{1.10}

\newcommand{\QBPerTokRatio}{1.54}

\newcommand{\QJacHi}{0.454}
\newcommand{\QJacLo}{0.066}
\newcommand{\QLfuEight}{36.0}
\newcommand{\QLruB}{75}
\newcommand{\QLruEight}{30.1}
\newcommand{\QStaticB}{73}
\newcommand{\QStaticEight}{33.9}
\newcommand{\QTopEightHi}{42}
\newcommand{\QTopEightLo}{34}
\newcommand{\QWOneHi}{45}
\newcommand{\QWOneLo}{38}

\newcommand{\QWSixtyHi}{96}
\newcommand{\QWSixtyLo}{95}
\newcommand{\ReplayEff}{29.6}

\newcommand{\RpCgOff}{16.5}
\newcommand{\RpCgOn}{16.5}

\newcommand{\RpMbOff}{18.1}
\newcommand{\RpMbOn}{16.4}

\newcommand{\SbBalLockPct}{81}
\newcommand{\SbBalOffDisk}{16.8}

\newcommand{\SbBalOnDisk}{33.6}

\newcommand{\SbBalOnScan}{2.4}
\newcommand{\SbBalVsCg}{2.0}
\newcommand{\SbNinetyLockPct}{74}
\newcommand{\SbNinetyOffDisk}{23.7}

\newcommand{\SbNinetyOnDisk}{18.5}

\newcommand{\StepBytesGB}{29.2}
\newcommand{\TcDiskA}{5.3}
\newcommand{\TcDiskC}{9.8}
\newcommand{\TcDiskD}{17.4}
\newcommand{\TcDiskE}{33.6}
\newcommand{\TcMedA}{0.99}
\newcommand{\TcMedC}{1.37}
\newcommand{\TcMedD}{1.95}
\newcommand{\TcMedE}{2.76}

\newcommand{\TcSpreadA}{0.5}
\newcommand{\TcSpreadC}{2.1}
\newcommand{\TcSpreadD}{4.0}
\newcommand{\TcSpreadE}{1.4}
\newcommand{\TieBandPct}{5.6}
\newcommand{\TokBytesGB}{27.8}

\newcommand{\WorkedMeasTokps}{1.01}
\newcommand{\XvAgreeWorst}{14.9}
\newcommand{\XvAgreeHit}{5.1}
\newcommand{\MgOnDiskLo}{31.1}
\newcommand{\MgOnDiskHi}{50.9}
\newcommand{\XvMemEffC}{339}
\newcommand{\XvMemInterpDev}{0.5}

\newcommand{\qwen}{Qwen3-30B-A3B\xspace}
\newcommand{\dsv}{DeepSeek-V3\xspace}

\begin{document}

\title{Who Should Own the Expert Cache? Kernel-Managed Tiering for Trillion-Parameter MoE Inference}

\author{Yuan Si}
\affiliation{%
  \institution{University of Waterloo}
  \city{Waterloo}
  \country{Canada}
}

\author{Yufeng Lin}
\affiliation{%
  \institution{Independent Researcher}
  \city{Macau}
  \country{China}
}

\author{Daming Li}
\affiliation{%
  \institution{Independent Researcher}
  \city{Mountain View}
  \country{USA}
}

\author{Jialu Zhang}
\authornote{Corresponding author.}
\affiliation{%
  \institution{University of Waterloo}
  \city{Waterloo}
  \country{Canada}
}
\begin{abstract}
Mixture-of-experts models whose expert pools exceed DRAM capacity
require a weight-residency tier.  Existing systems manage it in user
space with expert-granular placement, frequency-based admission,
and explicit pinning.  We evaluate whether the operating system page
cache can instead serve as the expert tier, using router traces from
three MoE models with 128 to 896 experts per layer; the trillion-parameter
production model's traces are replayed natively against its full
\PoolTB{}\,TB expert pool on GH200 hardware.  Capacity is
enforced by three independent mechanisms.

Iteration time varies smoothly with cache size (run-to-run spread
$\leq 4\%$), and device traffic follows the same trend.  Under severe
pressure the outcome depends on reclaim: device traffic rises above
miss demand only when
MGLRU, the tested kernels' default, is combined with balloon-style,
mostly \texttt{mlock}ed memory, a result reproduced on two
machines; cgroup
limits and \texttt{mem=} boots show no such behavior, so
balloon-based studies can overstate low-capacity device traffic by
about $2\times$.  At equal enforced memory, kernel recency
serves essentially the same demand as an oracle static-frequency
policy computed from the replay trace.  In the \texttt{pread}-based
replay the oracle-pinned arena stays
\EtwoTaxC--\EtwoTaxD$\times$ faster---a gap that is the cost of the
page-cache hit and reclaim path---but its static table degrades under domain shift
while recency remains stable.  At \PilotRecall\% measured recall,
router lookahead changes median time by \PfGainP\% when delivered as
kernel readahead advice; perfect one-layer advice gains \PfGain\% through the same
interface and nothing through blocking reads.  End-to-end at ample capacity,
enabling page-cache admission speeds steady decode by
\EsixLo{}--\EsixHi$\times$ in a production CUDA engine with
token-identical outputs.  These measurements favor kernel-managed
eviction, with model knowledge applied to admission and predictive
advice.
\end{abstract}

\maketitle

\section{Introduction}

A trillion-parameter mixture-of-experts model routes every token
through a small, input-dependent subset of its weights.  In the
production model studied here, an accepted token requires on average
\AccPerTok{} expert reads of \ExpertMB{}\,MB each, or
\TokBytesGB{}\,GB drawn from a \PoolTB{}\,TB pool.  No single node
holds that pool in DRAM, so serving systems in this regime require a
weight-residency tier.  Existing MoE serving systems typically manage
expert residency in user space using activation statistics, explicit
pinning, prefetching, and CPU/GPU or storage offload
\cite{moeinfinity,deepspeedmoe,mixtraloffload,sidamoe,pregated,fiddler}.
The production engine instrumented in this study bypasses the page
cache entirely (\S\ref{sec:background}).

We evaluate an alternative design: map the expert pool, read it through
the zero-copy path of a companion study \cite{ingestiontax}, and let
the kernel page cache manage host-memory residency.  We capture router
traces from three MoE models spanning 128 to 896 experts per layer.
The production model's traces are replayed natively against the full
\PoolTB{}\,TB pool on a GH200 node; the other two models drive
the fair-window policy simulations of \S\ref{sec:traces}, and
\qwen{} is additionally replayed on Darwin (\S\ref{sec:tc}).  The capacity
sweep uses three randomized runs per point, and the equal-memory
comparison two interleaved repetitions; every replay campaign uses
block-layer
I/O measurement and capacity control through a balloon, a cgroup wall,
or a physical \texttt{mem=} boot.  When reclaim policy is held
fixed, at least two capacity mechanisms give consistent results for
the capacity sweep of \S\ref{sec:tc}.  The equal-memory and prefetch experiments
(\S\ref{sec:pinning}, \S\ref{sec:advisor}) use the cgroup wall,
and \S\ref{sec:knee} analyzes the one configuration in which the
mechanisms disagree.  The static-frequency baseline is given oracle
access to the trace being replayed, which makes the comparison
favorable to that policy.

\begin{itemize}
\item \textbf{Capacity is a smooth, measurable parameter.}  Iteration
time varies smoothly with cache size (run-to-run spread $\leq$4\%),
and measured device traffic follows the same monotone trend.  A demand
model without fitted bandwidth constants compares simulated misses
with block-layer device counters and identifies re-read amplification
at each capacity (\S\ref{sec:tc}).
\item \textbf{Low-capacity amplification depends on reclaim configuration.}
Low-capacity amplification requires both MGLRU and a balloon-style
host with a high fraction of \texttt{mlock}ed memory.  It appears in
the balloon at \SbBalLockPct\% locked memory ($C{=}64$\,GB) and not at
\SbNinetyLockPct\% ($C{\approx}93$\,GB); the non-balloon
cells---the cgroup wall at $C{=}64$\,GB and the physical
\texttt{mem=} boot at ${\approx}93$\,GB---show no amplification.  The
result is reproduced on two machines and shows that balloon-based
sizing experiments can overstate low-capacity device traffic by
approximately $2\times$ (\S\ref{sec:knee}).
\item \textbf{Recency is effective at host-tier capacities.}  Under
identical evaluation windows, LRU matches or exceeds the same-domain
oracle static-frequency table from $B{=}32$ experts per layer upward,
which covers every host-memory budget evaluated here.  At trillion scale it
reaches approximately 70\% of Belady's optimum.  At the smallest
simulated budget ($B{=}8$), frequency performs better because LRU
cannot retain the immediate reuse set.  Under domain shift, the static
table falls to \ColMPin--\ColHPin\% while LRU remains at
\ColHLru--\ColCLru\%; global and per-layer allocation differ by at most
\GsGlobalGapMax{}\,pp (\S\ref{sec:traces}, \S\ref{sec:pinning}).
\item \textbf{Pinned ownership retains a bounded mechanism advantage.}
Under an enforced equal-memory limit with verified \texttt{mlock}
state and warm pages charged to the same budget, the oracle-pinned
arena is \EtwoTaxC--\EtwoTaxD$\times$ faster than the page cache in
the pread-based replay, with nearly identical device traffic.  The
device counters rule out byte selection as the cause; the residual
cost lies in the page-cache hit and reclaim path
(\S\ref{sec:pinning}).
\item \textbf{Delivery semantics limit the value of prediction.}  The
router predictor's recorded plan, replayed at \PilotRecall\%
measured recall, changes median time by \PfGainP\% when delivered
through \texttt{fadvise}, and yields no measurable benefit through
synchronous prefetch.  Perfect one-layer advice gains \PfGain\%
through the same advisory interface and nothing through blocking
reads.  Belady bounds the remaining miss-reduction potential at
roughly one third of the misses LRU still incurs (\S\ref{sec:advisor}).
\item \textbf{Measured constants give the placement rules.}
Admission follows from the miss already incurred, and the promotion
threshold from measured bandwidths, with no fitted parameters.  In a
production engine at ample capacity, enabling
page-cache admission improves steady decode by
\EsixLo--\EsixHi$\times$ with token-identical outputs
(\S\ref{sec:ownership}, \S\ref{sec:design}).
\end{itemize}

\section{Background and the Incumbent Design}
\label{sec:background}

\paragraph{The workload.}  A decoder-only MoE layer routes each token
to $k$ of $E$ experts.  The production model studied here uses
$k{=}16$ and $E{=}896$ across 92 routed layers, for $82{,}432$ experts
totaling \PoolTB{}\,TB at int4; its dense ``spine'' (attention,
shared projections, and routers) occupies a further 114\,GB
\cite{kimik3card}.  During decode, the spine is read at every step while
the routed subset changes.  The resulting expert
traffic---\ExpertsPerStep{} expert reads, \StepBytesGB{}\,GB per
engine iteration---is the workload studied in this paper.  Two open
models extend the expert-count range: \qwen{} ($k{=}8$, $E{=}128$, 48
layers, 17.6\,GB) and \dsv{} ($k{=}8$, $E{=}256$, 58 routed layers of
61) \cite{qwen3,deepseekv3}.  Assumptions that hold at 128 experts
per layer no longer hold at 896; the observed transition lies between
128 and 256 (\S\ref{sec:traces}).

\paragraph{The incumbent design.}  Systems for larger-than-memory MoE
inference commonly manage expert residency explicitly in user space,
using activation statistics, prefetching, pinning, and CPU/GPU or
storage offload
\cite{moeinfinity,deepspeedmoe,mixtraloffload,sidamoe,pregated,fiddler,edgemoe}.
This design entails three testable assumptions: expert popularity is
stable enough to guide placement (\S\ref{sec:traces}), the expert is an
appropriate allocation granularity (\S\ref{sec:pinning}), and admitting
expert pages to the OS cache would be counterproductive
(\S\ref{sec:pinning}, \S\ref{sec:design}).  The production engine used
for the traces applies the third assumption throughout:
\texttt{F\_NOCACHE} on every expert read.

\paragraph{Capturing routing.}  Routing is captured through a trace
hook in the production engine and a 40-line evaluation callback in
llama.cpp.  The protocol records only decode, because batched prefill
mixes positions and llama.cpp's scheduler dry run emits a synthetic
first pass, identifiable from bit-identical ``step~0'' coverage across
four prompts.  Replay uses the same per-expert files read by the engine,
so the replayed I/O stream corresponds directly to engine I/O.

\paragraph{What capacity denotes.}  We keep six quantities distinct:
the target provided to a capacity mechanism (``capacity $C$'');
\texttt{MemAvailable}; resident file-backed pages; locked anonymous
pages (\texttt{VmLck}); block-device reads from
\texttt{/proc/diskstats} sectors, which measure block-layer request
bytes rather than physical-media bytes; and logical expert accesses
from the trace.  Three independent mechanisms enforce $C$: a balloon,
a cgroup-v2 \texttt{MemoryMax} wall, and a boot-time \texttt{mem=}
limit.  The mechanisms agree except under the reclaim configuration
analyzed in \S\ref{sec:knee}.  Every run records
\texttt{MemAvailable}, \texttt{Cached}, \texttt{Mlocked}, and
\texttt{Unevictable} at the start of the measured loop, together with
the cgroup peak when applicable.

\paragraph{A canonical trace parser.}  Speculative decoding makes
``one token'' an ambiguous accounting unit.  An engine iteration may
contain two positions, represented by a 32-index record when
speculation succeeds, or may route one layer twice when a failed
speculation is followed by a fallback pass.  Collapsing these cases
biases downstream denominators.  Because the trace JSONL has no
explicit event tag, the canonical parser reconstructs each record as
\emph{normal}, \emph{spec-accept}, or \emph{spec-fallback} from record
width and per-layer duplication.  Across all four domains, the typed
accounting recovers exactly \AcceptedB{} accepted tokens per trace for
an engine request of 300.  Replay timing is reported in engine
iterations and serving arithmetic in accepted tokens.  All subsequent
analyses use this typed parse.

\paragraph{The read path.}  A companion study establishes a
three-condition execution contract under which file-backed weights
reach accelerator-class bandwidth without an intermediate
framework-owned copy: adopt the mapping inside the framework execution
path, keep activations accelerator-resident, and order work on the
accelerator queue \cite{ingestiontax}.  Under this contract, the GPU
reads a page-cache-resident expert at the same bandwidth as a
framework-owned allocation, so residency policy determines performance
once the copy path has been removed.  We use the execution contract
from that study and re-measure every bandwidth constant on the
evaluation node: \BwHbm{}\,GB/s for HBM weight reads;
\BwAdopt{}\,GB/s for GPU reads from file-backed host pages;
\BwHostAlloc{}\,GB/s for framework-owned \texttt{hostAlloc} storage;
\BwMalloc{}\,GB/s for unregistered \texttt{malloc} storage;
\DiskGBps{}\,GB/s for NVMe; and \BwStream{}\,GB/s for overlapped
copy-then-compute streaming over a link with a \BwLink{}\,GB/s
ceiling.  A second independent run reproduces each rate within
\BwRunDev{}\% (computed on unrounded rates).  This paper evaluates
which file-backed pages should remain resident and whether eviction
should be managed by the kernel or by a user-space expert cache.

\section{Router Locality at Three Scales}
\label{sec:traces}

Frequency-based pinning assumes that a sufficiently concentrated and
stable set of experts can be identified in advance.  We test this
assumption using per-layer top-$k$ router selections from four workload
domains---systems prose, code generation, mathematics, and travel
planning (labeled \emph{chat} in the figures and tables)---on the
three models in \S\ref{sec:background}
\cite{deepseekv3,qwen3,kimik3card}.  The production-model traces come
from its serving engine (242--290 engine iterations, \AcceptedB{}
accepted tokens per domain); \qwen{} and 671B \dsv{} are traced
through an evaluation callback added to llama.cpp \cite{llamacpp} (400
and 300 steps per domain, plus 4{,}000-step \qwen{} runs for coverage
asymptotics).  Prefill records are excluded throughout
(\S\ref{sec:background}).

\begin{table}[t]
\centering\small
\setlength{\tabcolsep}{1.9pt}
\begin{tabular}{lccc}
\toprule
 & Qwen3-30B & DeepSeek-V3 & Kimi-K3 \\
 & $E{=}128$ & $E{=}256$ & $E{=}896$ \\
\midrule
Zipf slope $\alpha$ & \QAlphaLo{}--\QAlphaHi{} & \DAlphaLo{}--\DAlphaHi{} & \KAlphaLo{}--\KAlphaHi{} \\
Reuse, $W_1$ & \QWOneLo{}--\QWOneHi{}\% & \DWOneLo{}--\DWOneHi{}\% & \KWOneLo{}--\KWOneHi{}\% \\
Reuse, $W_{64}$ & \QWSixtyLo{}--\QWSixtyHi{}\% & \DWSixtyLo{}--\DWSixtyHi{}\% & \KWSixtyLo{}--\KWSixtyHi{}\% \\
Domain Jaccard & \QJacLo{}--\QJacHi{} & \DJacLo{}--\DJacHi{} & \KJacLo{}--\KJacHi{} \\
LRU vs.\ static ($B{\geq}32$) & tie & LRU wins & LRU wins \\
\bottomrule
\end{tabular}
\caption{Decode-time routing locality across four domains and
three models; entries are min--max across the domains.  $W_n$ is the
fraction of selections repeated within the previous $n$ tokens;
Domain Jaccard is the overlap of per-domain top expert sets.}
\label{tab:locality}
\end{table}

\paragraph{Fact 1: concentration decreases with expert count.}
Rank--frequency slopes are Zipf-like in all three models: $\alpha$ is
\QAlphaLo{}--\QAlphaHi{} at $E{=}128$ and near or below 1 at the
larger pools (\DAlphaLo{}--\DAlphaHi{} at $E{=}256$,
\KAlphaLo{}--\KAlphaHi{} at $E{=}896$).  At trillion scale, the top-8
experts per layer carry only \KTopEightLo--\KTopEightHi\% of
selections, compared with \QTopEightLo--\QTopEightHi\% at $E{=}128$.

\paragraph{Fact 2: concentration varies sharply across layers.}  In the production
model, the first routed layer assigns only \DepthTopFirst{}\% of its
selections to the top eight experts, the most concentrated layer
reaches \DepthTopMax{}\%, a \DepthGradient$\times$ spread, and the
final layer sits at \DepthTopLast{}\%.  Per-layer union coverage
varies from \DepthUniqMax{} of 896 experts at the widest layer to
\DepthUniqMin{} at the narrowest, with early layers covering more of
the pool than the deepest quarter.  \dsv{} shows a similar profile.
The widest layers demand the most cache capacity, and a global LRU
pool adapts its allocation to this per-layer variation, whereas a
static per-layer quota must be tuned for each model.

\paragraph{Fact 3: at $E{=}128$, concentration increases in long
contexts.}  In the 4{,}000-step \qwen{} runs, the fitted slope rises to
\LongAlphaLo{}--\LongAlphaHi{}, and 64-token reuse rises from
\QWSixtyLo{}--\QWSixtyHi\% (Table~\ref{tab:locality}) to about 99\%.
The longer traces are therefore at least as cacheable as the shorter
ones.

\paragraph{Fact 4: recency is strong at every scale.}  At $W_1$,
\KWOneLo{}--\KWOneHi{}\% of the production model's selections repeat
the previous token's experts, rising to \KWSixtyLo{}--\KWSixtyHi{}\%
within 64 tokens.  Most of that reuse is already present at $W_8$
(\KWEightLo{}--\KWEightHi{}\%), so a modest cache captures much of the
available recency.

\paragraph{Fact 5: static hot sets do not transfer across domains.}
The top expert sets of different domains overlap by as little as
\QJacLo{} (Jaccard) in \qwen{} and at most \KJacHi{} in the production
model (Table~\ref{tab:locality}).  A pin set learned on one workload
therefore has limited overlap with the hot set of another.  Facts~1--4
show substantial locality, but that locality is expressed more
reliably through recent reuse than through a fixed global ranking.

\paragraph{Fact 6: coverage continues to grow without saturating.}
Within 300 steps, every domain touches \KCovLo--\KCovHi\% of the
production pool.  The 4{,}000-step \qwen{} runs show the longer trend:
coverage reaches \LongCovLo--\LongCovHi\% and continues to increase at
4{,}000 steps, while each additional step reuses recently
selected experts at approximately 99\%.  Because coverage does not saturate, no fixed
subset of the pool suffices: dropping experts changes the executed
model, whereas caching preserves the full pool and pays the measured
miss latency instead (\S\ref{sec:tc}).

\begin{figure}[t]
\centering
\includegraphics[width=\columnwidth]{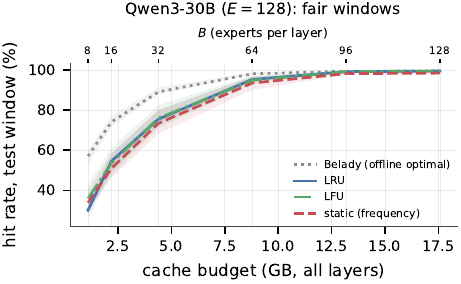}\\[2pt]
\includegraphics[width=\columnwidth]{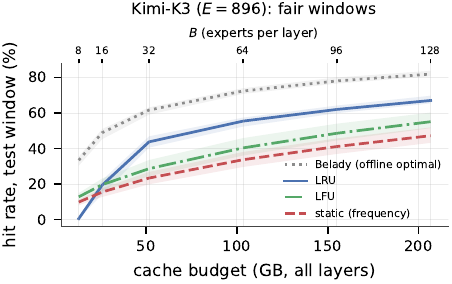}
\caption{Fair-window hit rates at the two ends of the expert-count
range (line: mean over four domains; band: per-domain min--max).
$E{=}256$ lies between them for $B{\geq}32$ (LRU \DLruB\% at
$B{=}32$); at $B{=}8$ it already favors LRU.}
\label{fig:policy}
\end{figure}

\paragraph{At the smallest budgets, recency thrashes.}
Fig.~\ref{fig:policy} evaluates static frequency pinning, LFU, LRU, and
Belady's offline optimum under identical windows: every policy warms on
the first 60\% of iterations and is scored on the final 40\%.
Comparing a warm frequency table against cold-start LRU produces an
artifactual crossover that disappears once
both policies use the same warm-up and scoring windows; the
small-budget inversion below survives fair windows.  At the
smallest simulated budget ($B{=}8$, at or below the routed fan-out
$k$), LRU evicts entries
before their next reuse.  At $B{=}8$ with $k{=}16$, it hits
\KLruEight\% while LFU reaches \KLfuEight\% and static frequency
\KStaticEight\%.  Frequency therefore performs best in this tiny-cache
regime, and the same inversion appears at $E{=}128$ (LRU \QLruEight\%
versus LFU \QLfuEight\% and static \QStaticEight\%).

\paragraph{At host-tier budgets, recency wins and remains ahead.}
From $B{=}32$ upward---approximately 52\,GB at production scale, just
below the smallest measured capacity ($C{=}64$\,GB;
Table~\ref{tab:tc})---the ranking reverses and remains stable.  At
$E{=}128$, the two policies fall within the domain spread (\QLruB\%
LRU versus \QStaticB\% static at $B{=}32$).  At $E{=}256$, LRU leads
at every simulated budget, including $B{=}8$, and reaches \DLruB\%
versus \DStaticB\% at $B{=}32$; at $E{=}896$ it leads from $B{=}16$
upward in the per-layer regime
(\KChatLruSixteen\% versus \KChatStaticSixteen\% on the chat domain
at $B{=}16$, and
\KLruB\% versus \KStaticB\% at $B{=}32$).  At trillion scale, LRU captures
approximately 70\% of Belady's optimum (\KLruB{} of \KBeladyB\%),
whereas static frequency captures roughly one third (\KStaticB{} of
\KBeladyB\%).  As the expert head becomes less concentrated and the hot
set shifts across domains, a static ranking becomes stale while recency
adapts.

\paragraph{Bounds in the policy comparison.}  In
Fig.~\ref{fig:policy}, the Belady bound falls
as $E$ grows---larger expert pools are intrinsically harder
to cache---and the LRU-to-Belady gap widens with $E$.  The
miss-reduction potential available to predictive advice
(\S\ref{sec:advisor}) therefore grows with the expert pool, whereas a
static rank list does not close the gap.

\paragraph{Global-pool and per-layer results agree on these traces.}
A page cache is one pool keyed by (layer, expert), rather than 92 fixed
per-layer caches.  Layered-paging theory shows that fixed per-layer
allocations can have an unbounded competitive ratio in the worst case
\cite{llru}.  We therefore simulate both allocation regimes on the
production model's prose trace and its evaluation window: one global
budget of $N$ experts shared
across layers and the equal-split per-layer budget used above.  On
that trace, we do not observe the worst-case separation: global and
equal-split LRU remain within \GsGlobalGapMax{}\,pp at every budget,
and the policy ordering is identical in both regimes at every budget
from $B{=}32$ upward.  At the 1{,}472-expert budget ($B{=}16$) the two
regimes disagree on LRU versus static (16.8\% versus 18.8\% global,
19.5\% versus 17.5\% per-layer).  The global pool also allocates fewer
distinct experts to the more concentrated layers without
model-specific quotas.

\paragraph{Batching increases simulated misses.}  Merging the four
domain traces iteration-by-iteration through the canonical parser
preserves every speculative and fallback pass; the union is truncated
to the 242 iterations of the shortest trace, which places \KBAccepted{}
of the $4\times\AcceptedB{}$ accepted tokens inside the merged window.
The batch-4 union transfers \KBGBStep{}\,GB per step, corresponding to
only \KBDedup$\times$ deduplication.  Four streams sharing one budget
lose hit rate under the same 60/40 window: at a 25\%-of-pool budget,
the rate falls from \KBCbTwoFiveOne{}\% for one stream to
\KBCbTwoFiveFour{}\% for batch 4.  The resulting rise in misses is
well below the $4\times$ demand of four independent streams, but still
yields a simulated per-token miss ratio of \KBPerTokRatio$\times$ at a
fixed ${\approx}200$\,GB budget ($B{=}\KBFixedB$ per layer); the
corresponding ratio for \qwen{} is \QBPerTokRatio$\times$ at $B{=}32$,
${\approx}4.4$\,GB.

\finding{routing}{At trillion scale, routing is relatively flat
($\alpha\!\approx\!1$), the top-8 experts carry only
\KTopEightLo--\KTopEightHi\% of selections, hot sets vary by domain,
and coverage continues to grow, while recency remains high
($W_{64}$=\KWSixtyLo--\KWSixtyHi\%).  The policy ordering changes
between $E{=}128$ and 256: from $B{=}32$ upward, LRU exceeds the static
frequency oracle; at $B{=}8$, at or below the per-token fan-out,
LRU thrashes.  Global and per-layer allocation agree
within \GsGlobalGapMax{}\,pp on the prose trace.}

\section{Iteration Time and Device Traffic versus DRAM Capacity}
\label{sec:tc}

We measure iteration time and block-layer device traffic across DRAM
capacities under three capacity mechanisms.  Iteration time varies
smoothly with capacity under every mechanism; one reclaim
configuration additionally inflates device traffic above the
simulated miss demand, and \S\ref{sec:knee} analyzes it separately.

\paragraph{Method.}  We replay the production model's prose trace
against the full \PoolTB{}\,TB pool on a GH200
node~\cite{gh200} (480\,GB Grace DRAM, 3.9\,TB NVMe; machine~A),
through a native harness that reads exactly the experts
selected by each recorded router pass.  A second GH200 node of the
same memory configuration but with an NVMe class approximately
3$\times$ faster (machine~B) is used for the mechanism
cross-validation
below and for the equal-memory, prefetch, and end-to-end experiments of
\S\S\ref{sec:pinning}--\ref{sec:design}.  The canonical parser yields \ItersB{} iterations and
\AcceptedB{} accepted tokens, with speculative and fallback passes
replayed as issued.  For the principal sweep, capacity $C$ is set by an
\texttt{mlock}ed balloon~\cite{waldspurger}.  For each capacity
point, we drop caches, verify the locked size, run one warm-up
replay, and measure a second.  The sweep is repeated three times in randomized capacity order.
Every pass snapshots \texttt{/proc/diskstats}, \texttt{/proc/vmstat}, and
pressure-stall accounting, so device bytes are measured at the block
layer.

\begin{algorithm}[t]
\small
\DontPrintSemicolon
\KwIn{plan $P$, mode $M$, capacity $C$, scope name $u$}
terminate any existing scope $u$; \texttt{sync}; drop caches;
snapshot $S_{\mathrm{pre}}$\;
start scope $u$: \texttt{MemoryMax}$=C{+}6$\,GiB,
\texttt{MemorySwapMax}$=0$\;
\textbf{in} $u$: $\mathrm{replay}(P,M)$ \tcp*{warm-up; abort if
\texttt{mlock} fails}
\textbf{in} $u$: snapshot diskstats/meminfo/\texttt{VmLck} at loop
entry \tcp*{preload excluded}
\textbf{in} $u$: $\mathrm{replay}(P,M)\rightarrow$ per-iteration
wall, bytes \tcp*{measured pass}
snapshot $S_{\mathrm{post}}$; record peak \texttt{memory.current}
and the kernel log\;
\textbf{retain the run iff} \texttt{VmLck} matches the request
$\wedge$ no OOM event $\wedge$ the scope stays alive throughout\;
\caption{One measurement cell under the cgroup wall.}
\label{alg:cell}
\end{algorithm}

Cells enforced with the cgroup wall follow Algorithm~\ref{alg:cell}:
warm-up and measurement execute in one accounting scope.  With
separate scopes, file pages from the terminated warm-up scope are
reparented without being charged to the measured scope, and a nominal
$C{=}256$\,GB measurement reproduces the 448\,GB point; the per-run
\texttt{memory.current} timeline exposes the miscount.

\begin{figure}[t]
\centering
\includegraphics[width=\columnwidth]{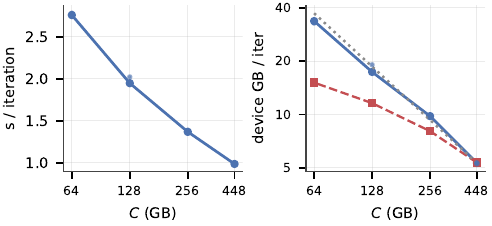}
\caption{Capacity response: iteration time (left, with all three
randomized runs) and block-layer device traffic (right, log--log).
Right panel: circles on a solid line, measured device reads, with the
three individual runs as faint circles; squares
on a dashed line, simulated nominal miss demand; dotted, the $a/C$
inverse-capacity reference anchored at the 448\,GB point.}
\label{fig:tc}
\end{figure}

\begin{table}[t]
\centering\small
\setlength{\tabcolsep}{2.9pt}
\begin{tabular}{rrrrrrr}
\toprule
$C$ (GB) & med (s) & spread & GB/it & $a/C$ dev & miss (GB) & amp \\
\midrule
448 & \TcMedA & \TcSpreadA\% & \TcDiskA & \McInvErrA\% & \McNomA & \McAmpA$\times$ \\
256 & \TcMedC & \TcSpreadC\% & \TcDiskC & \McInvErrC\% & \McNomC & \McAmpC$\times$ \\
128 & \TcMedD & \TcSpreadD\% & \TcDiskD & \McInvErrD\% & \McNomD & \McAmpD$\times$ \\
 64 & \TcMedE & \TcSpreadE\% & \TcDiskE & \McInvErrE\% & \McNomE & \McAmpE$\times$ \\
\bottomrule
\end{tabular}
\caption{The capacity sweep: three-run medians and iteration-time
spread; device traffic compared with the inverse-capacity reference
$a/C$ (anchored at the 448\,GB point) and with the constant-free demand
model (simulated miss GB per iteration).  The amplification column is
measured device bytes divided by nominal misses.  The 64\,GB row is
dominated by balloon-conditioned excess; the 256 and 128\,GB rows are
measured under a balloon and may carry a smaller component of it
(\S\ref{sec:knee}).}
\label{tab:tc}
\end{table}

\paragraph{The observed curve.}  Median iteration time increases
smoothly and monotonically from \TcMedA{}\,s at $C{=}448$\,GB to
\TcMedE{}\,s at 64\,GB, with run-to-run spread of
\TcSpreadA{}--\TcSpreadD{}\% (Table~\ref{tab:tc}).  The curve
covers one trace, one host, four capacities, and one kernel version.
Selected capacities are remeasured with one mechanism changed at a
time in \S\S\ref{sec:knee}--\ref{sec:pinning}, and
\S\ref{sec:ownership} sizes a deployment from it.

\paragraph{An inverse-capacity reference for device bytes.}  Measured
device traffic follows an inverse-capacity reference anchored at the
448\,GB point (Fig.~\ref{fig:tc}, right): halving the cache approximately
doubles NVMe traffic, from \TcDiskA{} to \TcDiskE{}\,GB per iteration,
with per-point deviations between \McInvErrE{}\% and \McInvErrC{}\%
(Table~\ref{tab:tc}).  The reference has no fitted parameter beyond its
anchor.

\paragraph{A constant-free demand diagnostic.}  The demand model does
not estimate elapsed time.  Raw-device probes read \ProbeCold{}\,GB/s
cold and \ProbeWarm{}\,GB/s warm, while the replay path's effective
rate is \ReplayEff{}\,GB/s---a spread of more than 2$\times$---so a
time model calibrated from any single bandwidth constant would not
transfer reliably across systems.  Instead, we simulate LRU at capacity
$C$, multiply misses by expert size, and compare the result with
measured device bytes (Table~\ref{tab:tc}).  A ratio above one
indicates re-read amplification; it rises from \McAmpA$\times$ at
448\,GB to \McAmpE$\times$ at $C{=}64$\,GB.  This diagnostic
identifies excess I/O without assigning a machine-specific time
constant.

\paragraph{Agreement across capacity mechanisms.}  Ballooning reserves
anonymous memory and treats the remainder as the capacity available to
the file tier.  We therefore cross-validate selected points under a
cgroup-v2 \texttt{MemoryMax} wall (Algorithm~\ref{alg:cell}) and under
\texttt{mem=} boots, in which the kernel cannot use the omitted DRAM.
When reclaim policy is held constant, the mechanisms agree in median
iteration time within \XvAgreeHit{}\% at the hit-dominated end (balloon
versus cgroup wall at $C{=}448$\,GB, across the two machines) and
within \XvAgreeWorst{}\% in the deepest matched cell ($C{=}64$\,GB with
MGLRU off, machine~A balloon versus machine~B cgroup wall).  A \texttt{mem=} boot at an effective
capacity of \XvMemEffC{}\,GB differs by \XvMemInterpDev{}\% from the
balloon curve interpolated to that capacity
(Fig.~\ref{fig:mglru}, right).  One further cross-machine pair differs by more than these figures
because the two machines' storage devices differ in bandwidth by
approximately 3$\times$; it is excluded from the agreement numbers.
Nominal $C$ is therefore used as the effective cache capacity.

\paragraph{Batched capacity.}  We replay the canonical batch-4 union
plan from \S\ref{sec:traces}, preserving all speculative and fallback
passes and accounting for \KBAccepted{} accepted tokens.
Per-accepted-token time is \EfivePerTokA{}\,s at $C{=}448$\,GB and
\EfivePerTokC{}\,s at 256\,GB, against \SingleTokA\ and
\SingleTokC{}\,s for the single-stream replay at the same capacities
(\EfiveVsA\% and \EfiveVsC\%).  The batch-4 replay's per-iteration
spread is also narrower than the single-stream spread at the same
capacity, consistent with the union of four router streams smoothing
aggregate miss demand.

\paragraph{Darwin response.}  An M-series Mac (128\,GB, Darwin, 16\,KB
pages) replaying \qwen{} also exhibits a smooth response, from
\MFullMs{}\,ms per step with the \MPoolGB{}\,GB pool fully cached to
\MSqueezeMs{}\,ms at a 9\,GB file cache.  Ballooning that leaves the
desktop responsive cannot reach the lowest-capacity Darwin regime, so
the reclaim attribution in \S\ref{sec:knee} is limited to Linux.  On Darwin, the
compressor absorbs balloon pressure before the file cache yields, and
Metal-wired pages are not evictable.

\finding{capacity}{Iteration time and device traffic vary smoothly with
capacity, with iteration-time run spread at most 4\%.  Capacity
mechanisms agree within \XvAgreeHit{}\% at the hit-dominated end and
\XvAgreeWorst{}\% in the deepest matched cell when reclaim policy is
held constant.  Batch-4 does not increase per-accepted-token time in
the replay.  The demand model identifies the low-capacity cells in
which re-read amplification requires separate reclaim analysis.}

\section{Low-Capacity Amplification Depends on Reclaim Configuration}
\label{sec:knee}

\begin{figure*}[t]
\centering
\includegraphics[width=\textwidth]{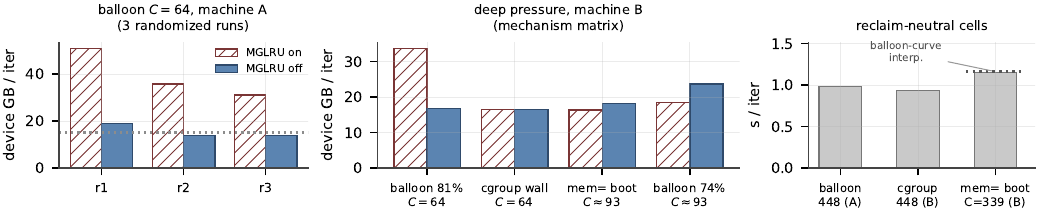}
\caption{Reclaim behavior.  In each paired group the left bar is MGLRU
enabled and the right bar MGLRU disabled.  Left: the three-run reclaim
toggle at $C{=}64$\,GB under balloon pressure on machine~A; the dotted
line is nominal miss demand.  Middle: the mechanism matrix on
machine~B---balloon at \SbBalLockPct\% locked and cgroup wall at
$C{=}64$\,GB, physical \texttt{mem=} boot and balloon at
\SbNinetyLockPct\% locked at an effective capacity of
${\approx}93$\,GB; amplification appears only in the
high-locked-fraction balloon.  Right (note the axis: seconds per
iteration, not device bytes): reclaim-neutral cells, all with
MGLRU enabled---balloon and cgroup wall at $C{=}448$\,GB and a
\texttt{mem=} boot at an effective capacity of \XvMemEffC{}\,GB, whose
dotted tick is the balloon iteration-time curve interpolated to that
capacity.}
\label{fig:mglru}
\end{figure*}

From $C{=}256$\,GB downward, the device traffic measured with the
balloon rises above the traffic predicted from misses
(\McAmpC$\times$ at 256\,GB, \McAmpD$\times$ at 128\,GB,
\McAmpE$\times$ at 64\,GB; Table~\ref{tab:tc}).  At $C{=}64$\,GB the
tier reads \TcDiskE{}\,GB per iteration against a nominal miss demand
of \McNomE{}\,GB, which implies that some pages are evicted and read
again within an iteration.  Toggling the reclaim implementation
isolates one contributing condition.  In a separate three-run reclaim
toggle on the same machine (distinct from the capacity sweep of
Table~\ref{tab:tc}, whose median for this cell is \TcDiskE{}\,GB),
MGLRU---the default in the tested
configuration---produces \MgOnDisk{}\,GB per iteration
(\MgOnDiskLo{}--\MgOnDiskHi{} across runs), with kswapd scanning
\MgOnScanLo{}--\MgOnScanHi{} pages per page reclaimed.  Classic LRU
produces \MgOffDisk{}\,GB per iteration
(\MgOffDiskLo{}--\MgOffDiskHi{}), with a scan-to-steal ratio of 1.0
(Fig.~\ref{fig:mglru}, left).  MGLRU is therefore necessary but not
sufficient for the amplification; the next comparison varies the
capacity mechanism as well.

\paragraph{Amplification requires two conditions.}  We repeat the
deep-pressure experiment on a second machine under all three capacity
mechanisms---the cgroup wall and the balloon at $C{=}64$\,GB, and
\texttt{mem=} at the lowest boot that leaves the system stable
(102\,GB visible, effective
capacity ${\approx}93$\,GB)---and toggle MGLRU within each condition
(Fig.~\ref{fig:mglru}, middle).  Under the cgroup
\texttt{MemoryMax} wall at $C{=}64$\,GB, traffic is effectively unchanged
between MGLRU enabled and disabled (\RpCgOn{} and \RpCgOff{}\,GB per
iteration).  Under the physical \texttt{mem=} boot,
MGLRU is also near nominal and slightly lower than classic LRU
(\RpMbOn{} versus \RpMbOff{}\,GB).  Under the balloon at $C{=}64$\,GB,
the same capacity as the cgroup wall, amplification returns: \SbBalOnDisk{}\,GB
with MGLRU and a scan-to-steal ratio of \SbBalOnScan{}, compared with
\SbBalOffDisk{}\,GB after disabling MGLRU.  Machine~A's reclaim
toggle, on a device
approximately 3$\times$ slower, records \MgOnDisk{}\,GB in the
corresponding cell.  The combined evidence attributes amplification to
the interaction between MGLRU and balloon pressure created by a very
large \texttt{mlock}ed anonymous region.  In the affected cell,
\SbBalLockPct\% of memory is locked.

\paragraph{Dependence on the locked fraction.}  Reducing the locked
fraction to \SbNinetyLockPct\% in the $C{\approx}93$\,GB balloon
condition removes the amplification on machine~B: MGLRU then
produces \SbNinetyOnDisk{}\,GB per iteration, compared with
\SbNinetyOffDisk{}\,GB under classic LRU.  Within the balloon
condition, amplification is therefore present at \SbBalLockPct\%
locked ($C{=}64$\,GB) and absent at \SbNinetyLockPct\% locked
($C{\approx}93$\,GB).  A balloon sets the locked fraction and the
remaining capacity with one knob, so this pair does not separate the
two variables by itself; the separation comes from the non-balloon
cells, which show no amplification at either capacity.  Classic LRU
under the balloon remains close to nominal in the median but has a
wider run-to-run range on machine~A
(\MgOffDiskLo--\MgOffDiskHi{}\,GB) than that of the cgroup-wall
repetitions, which differ by at most 0.1\,GB.  Ballooning is therefore a less stable
capacity-control mechanism even when it does not cause amplification.

\paragraph{Methodological consequence.}  Ballooning is widely used to
emulate smaller-memory machines~\cite{waldspurger}, and the principal
capacity sweep uses balloons with the tested kernel default, so its
lowest-capacity points include the additional traffic produced by the
combination of MGLRU and a high locked fraction.  The overstatement is
approximately 2$\times$: \SbBalVsCg$\times$ relative to the cgroup
wall at matched capacity on machine~B, and \MgAmp$\times$ between
MGLRU enabled and disabled on machine~A.  At an effective capacity of
approximately 93\,GB, the physical-memory condition records
\RpMbOn{}\,GB per iteration under \texttt{mem=}, whereas interpolation
from the balloon-conditioned curve implies approximately 24\,GB.
Table~\ref{tab:tc} therefore labels the rows below 448\,GB
balloon-conditioned, and physical-memory reductions are sized from the
\texttt{mem=} points.  Capacity studies that use ballooning
should cross-check at least one low-capacity point with a cgroup limit
or a \texttt{mem=} boot.  A generation-counter analysis of how MGLRU
ages pages behind a large unevictable region remains future work
\cite{mglru}.

\paragraph{Operator guidance.}  Production systems can create a similar
condition when large arenas are \texttt{mlock}ed for pinned expert tiers,
HBM staging pools, or other reservations.  On hosts with a large
unevictable fraction, operators can compare kswapd scan-to-steal ratios
with 1.0 and block-device bytes with nominal miss demand.  In our
measurements, a scan-to-steal ratio above 2 appears in every cell in
which amplification exceeds 2$\times$ and in no reclaim-neutral cell;
this is an empirical indicator for the tested systems, not a general
threshold.  The experiments support three mitigations: confining the
streaming reader in a memory-limited cgroup, reducing the locked
region, or benchmarking with MGLRU disabled.  The
last change reduces traffic by a factor of \MgAmp{} in the affected
condition.  Pressure-stall accounting~\cite{tmo} provides an additional
operational signal.

\finding{knee}{Low-capacity amplification requires both MGLRU and
balloon pressure with a very high locked fraction: it appears at
\SbBalLockPct\% locked ($C{=}64$\,GB) and is absent at
\SbNinetyLockPct\% locked ($C{\approx}93$\,GB), while the
non-balloon cells---the cgroup wall at $C{=}64$\,GB and the
\texttt{mem=} boot at ${\approx}93$\,GB---show no amplification.  In the
tested systems, balloon-based capacity measurements can overstate
low-capacity device traffic by approximately 2$\times$, and a kswapd scan-to-steal ratio above 2 identifies every cell in
which amplification exceeds 2$\times$.}

\section{Pinned Arena versus Page Cache at Equal Memory}
\label{sec:pinning}

\begin{figure}[t]
\centering
\includegraphics[width=\columnwidth]{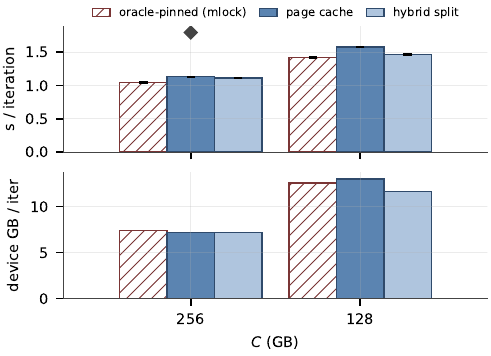}
\caption{Equal-memory comparison under one cgroup wall: iteration time
(top; error bars span the two interleaved repetitions) and
measured-interval device traffic (bottom).  The pinned and
hybrid arenas' one-time preloads ($C$ and $C/2$\,GB) are excluded from
the measured interval; \texttt{VmLck} verifies that the requested pages
are locked.  The diamond is the page cache with
\texttt{POSIX\_FADV\_RANDOM} at $C{=}256$\,GB.}
\label{fig:e2}
\end{figure}

User-space expert caches commonly pin frequently selected experts.  We
evaluate the mechanism and policy separately: whether serving a hit from
an \texttt{mlock}ed arena is faster than serving it from the page cache,
and whether a static frequency table retains better bytes than kernel
recency.  The comparison enforces equal memory for every run.

\paragraph{Setup.}  All configurations use the same native harness,
thread pool, and destination buffers, and perform one copy per access;
only the source of bytes on a hit differs.  The \emph{pinned} configuration
preloads the most frequent experts into one \texttt{mlock}ed arena.  Its
frequency table is computed from the evaluated trace, so it is an oracle
upper bound rather than a deployable policy.  Misses read from the full
pool.  The \emph{page-cache} configuration receives the same budget,
held entirely as file cache and with no policy tuning, and the
\emph{hybrid} configuration divides the budget between an arena and the
page cache.  Equal memory is enforced with the cgroup protocol in
Algorithm~\ref{alg:cell}.  At 256\,GB, retained pinned runs
record \texttt{VmLck}=\EtwoVmLckC{}\,GiB and \EtwoPinPreC{}\,GB of
one-time preload traffic outside the measured interval, and every
retained run reaches its peak \texttt{memory.current} at the wall with no
OOM event.  We run two interleaved repetitions per point in randomized
order.

\paragraph{Mechanism overhead.}  At $C{=}256$\,GB, the oracle-pinned arena
completes an iteration in \EtwoPinC{}\,s, compared with
\EtwoPcC{}\,s for the page cache, a \EtwoTaxC$\times$ difference
(\EtwoTaxD$\times$ at 128\,GB; Fig.~\ref{fig:e2}).  Both configurations
read nearly the same amount from storage, \EtwoPinDiskC{} and
\EtwoPcDiskC{}\,GB per iteration; the device counters therefore rule
out byte selection as the cause, and the timing difference reflects the
page-cache hit path---lookup and mapping on hits, plus reclaim work at
the memory wall.  Disabling readahead with
\texttt{POSIX\_FADV\_RANDOM} increases the page-cache time to
\EtwoBRand{}\,s, so readahead is not the source of the gap.  The gap
is specific to this \texttt{pread}-based replay.  Under the
zero-copy execution contract~\cite{ingestiontax}, the GPU reads
resident file-backed and framework-owned pages at
\BwAdopt{} and \BwHostAlloc{}\,GB/s (\S\ref{sec:background}), so the
corresponding overhead there should be smaller than
\EtwoTaxC$\times$; that path is not timed per token in
this study.  A 50/50 hybrid split at $C{=}256$\,GB takes \EtwoHyC{}\,s,
between the two pure configurations.

\begin{figure}[t]
\centering
\includegraphics[width=\columnwidth]{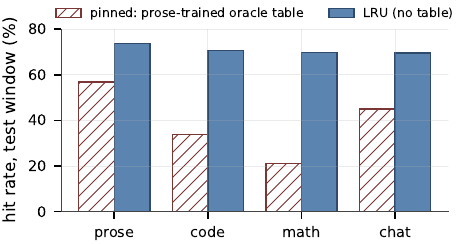}
\caption{Off-domain fair-window simulation of a prose-trained
oracle-frequency table and table-free LRU at a budget equivalent to
$C{=}256$\,GB ($B{=}159$ experts per layer), using the same
evaluation windows for both policies.}
\label{fig:collapse}
\end{figure}

\paragraph{Placement policy.}  We compare the fraction of total demand
served without device access.  At $C{=}256$\,GB the page cache serves
\EtwoPcSvcC\% and the oracle-pinned arena, on its own domain,
\EtwoPinSvcC\%
(\EtwoPinHitC\% from arena hits plus 1.4 points from misses served
out of the ${\approx}6$\,GiB of scope headroom above the arena;
Algorithm~\ref{alg:cell}); at $C{=}128$\,GB the
figures are \EtwoPcSvcD\% and \EtwoPinSvcD\%.  At equal capacity,
kernel recency therefore serves demand within about 1.6 points of an
oracle frequency table on the domain from which that table was
computed.  (The arena's table is built from the whole replayed trace;
the fair-window table below is trained on the first 60\% only, hence
its lower same-domain figure.)  Under domain shift the table degrades: in the fair-window
simulation of Fig.~\ref{fig:collapse} ($B{=}159$ experts per layer,
${\approx}256$\,GB), the prose-trained table reaches \ColPPin\% on its
own domain against LRU's \ColPLru\%, and falls to
\ColMPin--\ColHPin\% off domain while table-free LRU holds
\ColHLru--\ColCLru\%.  The trace analysis in \S\ref{sec:traces}
attributes this difference to cross-domain changes in expert selection,
with Jaccard similarity as low as \KJacLo.

\finding{policy}{At equal enforced memory, the oracle-pinned arena
is \EtwoTaxC--\EtwoTaxD$\times$ faster in the \texttt{pread}-based
replay, while serving essentially the same demand: page cache versus
arena is \EtwoPcSvcC\% versus \EtwoPinSvcC\% at $C{=}256$\,GB and
\EtwoPcSvcD\% versus \EtwoPinSvcD\% at 128\,GB.  The oracle table
loses hit rate under domain shift, whereas table-free LRU holds its
cross-domain hit rate.}

\section{The Router as a Prefetch Advisor}
\label{sec:advisor}

\begin{figure}[t]
\centering
\includegraphics[width=\columnwidth]{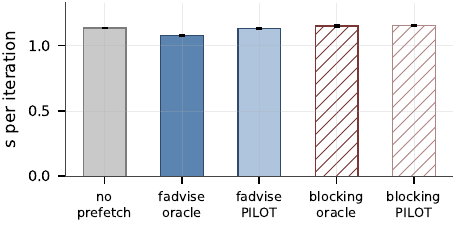}
\caption{One bounded prefetch worker under five conditions: no
prefetch; oracle next-pass hints delivered as advice or blocking reads;
and the measured PILOT prediction plan delivered through the same two
interfaces.  Error bars span the two
repetitions.  Medians are \PfNone{}\,s (none), \PfAdv{}\,s (oracle
advice), \PfAdvP{}\,s (PILOT advice), and \PfBlk{}/\PfBlkP{}\,s
(blocking oracle/PILOT).}
\label{fig:prefetch}
\end{figure}

Once layer $\ell$ has routed, a predictor can estimate the experts that
layer $\ell{+}1$ will require before their reads begin.  We measure both
the quality of this information and the interface used to deliver it.
Prediction quality and replay use one recorded expert-set plan: the
production engine writes it, and the replay reads it unchanged.  The
oracle next-pass
plan provides a one-layer upper bound.  Both plans use the same bounded
worker with one thread and one overwritable slot.  The worker either
submits \texttt{posix\_fadvise(WILLNEED)} or performs a blocking
\texttt{pread}, and it counts submitted, overwritten, and completed
hints in every condition.

\paragraph{Measured prediction quality.}  The production engine's
one-layer lookahead predictor (PILOT) caches next-layer gate weights
and recalls \PilotRecall\% of decode-time selections, with a per-layer
range of \PilotLayerLo{}--\PilotLayerHi\%.  Recall varies substantially
across layers, indicating that a uniform prefetch policy spends effort on
layers with limited predictability.  Previous-token recurrence, which
is available without model-specific lookahead, is \KWOneLo--\KWOneHi\%;
the learned predictor approximately doubles this baseline.

\paragraph{Delivery interface.}  At $C{=}256$\,GB under the cgroup wall,
oracle advice changes median iteration time from \PfNone{} to
\PfAdv{}\,s, an improvement of \PfGain\%; the oracle plan covers the
full next pass and submits 27{,}407 hints, none overwritten.  The
measured PILOT plan submits \PfSubm{} hints, none overwritten, and
yields \PfAdvP{}\,s, a \PfGainP\% change at \PilotRecall\%
recall---the same magnitude as the ${\approx}0.3$\% two-run spread.  The oracle and PILOT plans differ in hint volume as
well as recall, so their gap reflects both prediction quality and the
number of experts each plan covers.  Delivering the same plans through
blocking reads yields \PfBlk{}\,s for the oracle and \PfBlkP{}\,s for
PILOT, 1.5\% and 1.7\% slower than no prefetch
(Fig.~\ref{fig:prefetch}): a single bounded blocking worker does not
keep ahead of the eight-thread demand stream, and most of its reads
are overwritten in the slot before completion (17{,}553 of 27{,}407 for the
oracle plan).  The advisory interface issues readahead without delaying
demand and discards stale work when the slot is overwritten.

\paragraph{Lookahead depth and targeting.}  One layer provides a
natural horizon because the transfer can overlap with computation in the
current layer.  Deeper prediction compounds errors across layers and
reduces the probability that prefetched experts are used.  The observed
per-layer recall range motivates suppressing hints for poorly predicted
layers.  Such targeting is possible because, under advisory delivery,
an incorrect or late hint costs transfer efficiency without
changing the executed experts.  Belady's \KBeladyB\% hit rate compared with
LRU's \KLruB\% at $B{=}32$ shows that better future information could
eliminate approximately one third of the remaining LRU misses.

\finding{advice}{At \PilotRecall\% recall, the measured prediction plan
changes median iteration time by \PfGainP\%, within the two-run spread,
while oracle one-layer advice improves it by \PfGain\%.  Executing
either plan through a bounded blocking reader is slower than no
prefetch.  The appropriate integration point is
nonblocking advisory prefetch, with better prediction and
layer-specific targeting required to approach the Belady headroom.}

\section{Placement Rules and the Serving Procedure}
\label{sec:ownership}

The capacity response in \S\ref{sec:tc} characterizes the DRAM tier.  A
complete serving node also includes accelerator memory and storage.
Combining the bandwidth constants re-measured on the evaluation node
(\S\ref{sec:background}) with that capacity response yields three
placement rules and, with them, the serving procedure of
Algorithm~\ref{alg:serve}.

\paragraph{Rule 1: admit demand-read pages to DRAM}  A miss already
incurs the NVMe read, and allowing the resulting file pages to remain
in the page cache adds no I/O to that miss.  The remaining risk is
that admitted pages displace a more valuable working set;
Algorithm~\ref{alg:serve} removes that risk directly by keeping the
dense spine accelerator-resident or separately locked.  Retaining
demand-read expert pages then preserves the
possibility of future hits and yields the \AbSpeedup$\times$ repeated-read
improvement measured in \S\ref{sec:design}.

\paragraph{Rule 2: promote high-reuse experts to HBM}  Promoting an
expert of $S$ bytes to HBM costs one overlapped transfer
$S/\BwStream$ and saves $S(1/\BwAdopt - 1/\BwHbm)$ on each subsequent
use, with the rates in GB/s.  The breakeven reuse count is
\[
r^\ast \;=\; \frac{1/\BwStream}{1/\BwAdopt - 1/\BwHbm}
\;=\; \KStar.
\]
On the measured coherent link, the first reuse repays the transfer.
HBM allocation can therefore prioritize experts with the highest
expected reuse until capacity is exhausted.

\paragraph{Rule 3: prefer direct file-backed reads when adoption and
streaming have similar rates.}  Direct GPU reads from file-backed host
memory are within \TieBandPct{}\% of overlapped streaming on the measured
node (computed on unrounded rates).  In this range, file-backed adoption
avoids the copy engine, staging buffers, and double-buffering required
by explicit streaming.

\begin{algorithm}[t]
\small
\DontPrintSemicolon
\texttt{mmap} the pool; remove unbuffered/drop-behind flags from
expert reads \tcp*{\AbSpeedup$\times$ on repeated reads}
keep the dense spine accelerator-resident or separately locked
\tcp*{admitted experts cannot displace it}
\ForEach{decode layer $\ell$}{
  read $E_\ell$ through the page cache on demand\;
  after $\mathrm{router}(\ell)$ predicts
  $\widehat{E}_{\ell+1}$:
  \texttt{fadvise(WILLNEED,} $\widehat{E}_{\ell+1}$\texttt{)}
  \tcp*{advisory only; do not block demand}
}
export kswapd scan/steal and block-device bytes; alert when measured
reads exceed computed misses \tcp*{\S\ref{sec:knee}}
\lIf{the host has a large unevictable region}{confine the reader in a
memory-limited cgroup}
\lIf{the domain mix is stable}{optionally add an
\texttt{mlock}ed tier (\EtwoTaxC--\EtwoTaxD$\times$ in the
\texttt{pread} replay; smaller on the zero-copy path)}
\caption{Kernel-tier expert serving.  The kernel manages eviction;
the engine supplies model-specific admission and predictive advice.}
\label{alg:serve}
\end{algorithm}

\paragraph{Dependence on memory topology.}  The relative value of these
rules changes with the accelerator-memory topology~\cite{ingestiontax}.
On unified-memory systems, adopted file-backed pages and
framework-owned allocations occupy the same physical memory domain, so
host capacity and reclaim policy dominate.  On discrete PCIe systems,
unregistered host reads are limited to the \BwMalloc{}\,GB/s regime in
our measurements, making HBM ownership or explicit streaming necessary
for high bandwidth.  Coherent-link systems such as the measured GH200
support all three paths: file-backed host reads, HBM promotion, and
streaming.  The kernel-owned DRAM tier is therefore most directly useful
on unified- and coherent-memory systems; on PCIe systems, it remains a
storage tier below an HBM-resident working set.

\paragraph{Worked sizing example.}  The same measurements support
capacity planning in the style of the five-minute
rule~\cite{fiveminute}.
For a target maximum iteration time of two seconds, Fig.~\ref{fig:tc}
indicates approximately 128\,GB of DRAM, where the measured median is
\TcMedD{}\,s and device traffic is \TcDiskD{}\,GB per iteration.  A
256\,GB allocation yields \TcMedC{}\,s at \TcDiskC{}\,GB.  These
rows are balloon-conditioned (\S\ref{sec:knee}); treat the 128\,GB
figure as an upper bound on the DRAM the target requires.  At
$C{=}448$\,GB, the replay sustains \WorkedMeasTokps{} engine iterations
per second per stream, and the batch-4 replay requires \EfivePerTokA{}\,s
per accepted token; compute is excluded from these figures.  The
estimates apply to the measured replay, subject to the reclaim
conditions of \S\ref{sec:knee} and the batching behavior of
\S\ref{sec:tc}.

\finding{tiers}{Demand-read admission adds no I/O to the miss that
loaded the page.  On the measured coherent link, HBM promotion breaks
even after $r^\ast{=}\KStar{}$ reuse, and direct file-backed reads are
within \TieBandPct{}\% of overlapped streaming.  Together, the
measured rates decide when to retain pages in DRAM, promote experts to
HBM, or fetch from storage.}

\section{End-to-End Validation in the Production Engine}
\label{sec:design}

Algorithm~\ref{alg:serve} requires no kernel modification.  We
evaluate its admission steps in the production engine: first in the
existing reader, then along the complete decode path.

\paragraph{Cache-admission microbenchmark.}  We use the production
engine's existing \texttt{pread} reader, including its threads, buffers,
and files, to repeat reads of an 11.2\,GB expert set.  Under the shipped
\texttt{F\_NOCACHE} policy, every pass incurs device traffic and reaches
\AbNocache{}\,GB/s.  Enabling cache admission raises throughput to
\AbCached{}\,GB/s, a \AbSpeedup$\times$ improvement under the same
reader and engine.

\paragraph{End-to-end validation.}  We run full decode through the
production engine on the GH200 with three measured pairs per prompt,
ordered admission-first, \texttt{F\_NOCACHE}-first, and by a seeded
shuffle.  A run is retained only if the engine reports the expected CUDA MoE
backend and produces a non-degenerate output.  At ample
capacity---no capacity mechanism applied, on the full 480\,GB
host---enabling
admission improves median steady-decode time by
\EsixLo{}--\EsixHi$\times$ per prompt (Table~\ref{tab:e2e}).  Total
time, which includes the
policy-independent cold model load, improves by
\EsixTotLo{}--\EsixTotHi$\times$.  The two runs
in each pair produce identical 60-token ID sequences and the same
SHA-256 output hash.  Table~\ref{tab:e2e} reports the median and
per-prompt range across the three pairs per prompt.  These runs are
single-stream; \S\ref{sec:tc} reports the batch-4 union replay.

\begin{table}[t]
\centering\small
\setlength{\tabcolsep}{4pt}
\begin{tabular}{lrrcc}
\toprule
prompt & admit (s) & \texttt{F\_NOCACHE} (s) & speedup [range] & pairs \\
\midrule
prose & \EsixKP & \EsixDP & \EsixXP$\times$ [\EsixXLoP--\EsixXHiP] & \EsixPairsP \\
code  & \EsixKC & \EsixDC & \EsixXC$\times$ [\EsixXLoC--\EsixXHiC] & \EsixPairsC \\
chat  & \EsixKH & \EsixDH & \EsixXH$\times$ [\EsixXLoH--\EsixXHiH] & \EsixPairsH \\
\bottomrule
\end{tabular}
\caption{End-to-end decode through the production engine: per-prompt
median steady-decode time (total time minus cold model load) with
page-cache admission enabled (\emph{admit}) and under the shipped
\texttt{F\_NOCACHE} reader, over three balanced-order pairs from three
of the four trace domains (mathematics was not run end to end).  The
speedup column is the median of the three per-pair ratios, not the
quotient of the two median columns.  Every pair has an identical output
hash.}
\label{tab:e2e}
\end{table}

\paragraph{Cross-process sharing.}  The page cache is system-wide, so
processes share one physical copy, benefit from pages loaded by another
process, and retain cached pages across process
restarts~\cite{ingestiontax}.  A conventional user-space arena is
private to each process, which can require $n$ copies of the hot set
for $n$ serving processes.

\paragraph{Validation scope.}  The end-to-end validation uses one
production engine.  On the 480\,GB host, the available open model
(17.6\,GB of experts in an 18.6\,GB file) does not exercise this tier:
llama.cpp's \texttt{--cpu-moe} places experts in anonymous memory and
bypasses the page cache, while without that option the model fits in
HBM.  A separate Apple-silicon experiment
measures decode time under mmap-based and buffered loading of the
same model and finds the two within \EngSpread\%.

\finding{engines}{Enabling cache admission improves repeated-reader
throughput by \AbSpeedup$\times$ and median steady decode by
\EsixLo--\EsixHi$\times$ on the validated CUDA path at ample
capacity, with identical output-token sequences across paired runs.}

\section{Limitations}
\label{sec:validity}

\paragraph{Replay scope.}  The capacity, policy, and prefetch campaigns
replay recorded router passes through the expert I/O tier with compute
excluded.  The end-to-end measurements in \S\ref{sec:design} evaluate
how this tier composes with a production engine at ample capacity.
The replay results therefore characterize I/O-tier behavior and provide
bounds for, rather than complete measurements of, deployed serving.

\paragraph{Oracle upper bound.}  The pinned arena's frequency table is
computed from the evaluated trace.  This gives the arena information
that a deployed system would not possess and biases the comparison in
its favor.  The off-domain experiment in
Fig.~\ref{fig:collapse} directly measures the sensitivity of the static
table to workload change.

\paragraph{Scope of the reclaim attribution.}  The reclaim experiments
cover two machines, Linux 6.8, and one NVMe device class per machine.
The attribution to the interaction between MGLRU and high-locked-fraction
balloon pressure is behavioral: it rests on same-machine comparisons
across three capacity mechanisms and two locked fractions on
machine~B, with the balloon reclaim toggle reproduced independently on
machine~A and the MGLRU state recorded for every run.  A
generation-counter analysis explaining how pages age behind the large
unevictable region remains future work.  Balloon traffic is also
unstable from run to run in both reclaim conditions: with MGLRU enabled it spans
\MgOnDiskLo--\MgOnDiskHi{}\,GB per iteration across repetitions, and
with MGLRU disabled \MgOffDiskLo--\MgOffDiskHi{}\,GB, a
\MgOffSpread\% range, so the disabled condition is close to nominal but
retains measurable variability.  Darwin reproduces smooth capacity
degradation but cannot safely reach the corresponding lowest-capacity
regime.

\paragraph{Traces.}  Production-model traces contain 242--290 engine
iterations and \AcceptedB{} accepted tokens per domain, with speculative
and fallback passes replayed as issued.  Four prompt families represent
the workload domains.  The 4{,}000-step asymptotic experiment uses the
128-expert model, so longer traces from the production model remain
future work.  Greedy decoding makes the recorded access streams
reproducible; the results do not establish behavior under stochastic
sampling.  Cross-domain results are reported as observed ranges across the four
workloads.

\paragraph{Correctness invariance.}  Under greedy decoding, the
recorded expert selections are the exact access stream issued by the
engine and are replayed in their original order; residency policy
changes only where the selected bytes are served from
(\S\ref{sec:design} reports the paired output hashes).

\paragraph{Scope exclusions.}  Multi-node serving, quality effects of
pruning instead of caching, and request-arrival processes are outside the
measurement scope.  Each adds a system or modeling dimension beyond the
single-node residency tier characterized here.

\paragraph{Reproducibility.}  The replay harness, simulators, and
analysis pipeline will be submitted for artifact evaluation.

\section{Related Work}
\label{sec:related}

\paragraph{Mixture-of-experts models.}  Sparse expert
layers~\cite{shazeer2017moe} scaled through GShard~\cite{gshard},
Switch~\cite{switch}, and GLaM~\cite{glam} to open models including
Mixtral~\cite{mixtral}, \qwen{}~\cite{qwen3},
\dsv{}~\cite{deepseekv3}, and the Kimi K2/K3
line~\cite{kimik2,kimik3card}.  These models
increase total expert capacity while activating only a subset per token.
The trace study in \S\ref{sec:traces} examines how increasing expert
count changes the relative performance of frequency and recency policies.

\paragraph{MoE offload and expert caching.}
MoE-Infinity~\cite{moeinfinity} uses observed routing behavior to prefetch
and cache experts in user space.  PowerInfer~\cite{powerinfer} places
frequently activated neurons on the GPU using offline profiles.  LLM in
a Flash~\cite{llmflash} reduces flash traffic through activation
windowing and increases transfer granularity through row-column
bundling.  DeepSpeed-MoE~\cite{deepspeedmoe} develops system and model
optimizations for large-scale MoE inference.  Mixtral-offloading uses an
expert LRU~\cite{mixtraloffload}; SiDA-MoE~\cite{sidamoe} predicts expert activation offline with a
hash-based sparsity model to place experts between GPU and host
memory; Pre-gated
MoE~\cite{pregated} moves gating ahead of expert transfer; Fiddler
\cite{fiddler} coordinates CPU and GPU expert execution; and
EdgeMoE~\cite{edgemoe} targets sparse-model inference on mobile devices.
FlexGen~\cite{flexgen} schedules dense-model offload and addresses a
complementary problem.

FineMoE uses
fine-grained expert-selection patterns and prompt-semantic hints to guide
prefetching, caching, and offloading~\cite{finemoe}.  SpecPrefetch uses a
lightweight predictor only to initiate asynchronous transfer, while the
native router determines the experts that execute~\cite{specprefetch};
this separation corresponds to the advisory interface evaluated in
\S\ref{sec:advisor}.  Angelopoulos et al.\ formulate expert caching as a
layered paging problem, show that fixed per-layer cache allocations can
have an unbounded competitive ratio, and propose a layer-aware extension of
LRU~\cite{llru}.  Motivated by that result, we compare equal per-layer
allocation with one global capacity pool in \S\ref{sec:traces} and find
a difference of at most \GsGlobalGapMax{}\,pp on the
production prose trace.

\paragraph{Kernel-owned expert tier.}  Prior expert-offload systems
primarily manage residency in user space, with different expert
granularities, prediction sources, and transfer mechanisms.  This paper
adds an equal-memory page-cache baseline (\S\ref{sec:pinning}),
measures the reclaim behavior inherited from the operating system
(\S\S\ref{sec:tc}--\ref{sec:knee}), and identifies page-granular
sharing across processes as a property a per-process arena cannot
provide (\S\ref{sec:design}).  A companion study establishes the zero-copy execution contract used
for the reads themselves~\cite{ingestiontax}
(\S\ref{sec:background}); the present paper evaluates residency
management for the pages being read.

\paragraph{LLM serving systems.}  PagedAttention~\cite{vllm} applies
paging to the KV cache in user space, and Orca~\cite{orca} introduces
iteration-level continuous batching.  Both concern activation memory and
scheduling and can be combined with a kernel-managed weight tier.
llama.cpp~\cite{llamacpp} uses mmap-based model loading by default; this
paper characterizes the capacity and workload conditions under which
that approach is effective for expert weights.

\paragraph{OS memory management.}  Ballooning as a capacity-control
mechanism descends from VMware ESX~\cite{waldspurger}.  TMO uses
pressure-stall accounting to manage datacenter memory
offload~\cite{tmo}, and TPP places pages across CXL memory tiers
\cite{tpp}.  The capacity-response curve in \S\ref{sec:tc} describes the
expert-tier demand that such controllers would need to satisfy.
MGLRU~\cite{mglru} is the reclaim implementation evaluated in
\S\ref{sec:knee}.  Database systems have documented limitations of
mmap-based buffer management~\cite{crotty}.  The evaluated expert pool is
read-only, sequential within each expert, and shareable across
processes, unlike the mutable transactional workloads those
limitations concern.

\paragraph{Caching theory.}  Prior caching work spans Belady's
optimal replacement~\cite{belady}, working sets
\cite{denning}, and policies combining recency and frequency, such as
ARC~\cite{arc}.  The trace experiments compare recency, frequency, and
Belady's optimum on MoE routing traces (\S\ref{sec:traces}).  The
five-minute rule~\cite{fiveminute} motivates the
reuse-based placement calculation in \S\ref{sec:ownership}.  Work on
caching with machine-learned advice~\cite{lykouris,rohatgi,antoniadis}
bounds the value of predictions; \S\ref{sec:advisor} measures how much
of that bound a one-layer predictor realizes through advisory and
blocking interfaces.

\section{Conclusion}

A \PoolTB{}\,TB MoE model cannot hold its expert pool in DRAM, so the
serving system must manage partial residency.  In the measured workloads,
a same-domain oracle frequency table at best matches recency, and its
hit rate
degrades under domain shift.  At equal enforced memory, an oracle-pinned
arena retains a \EtwoTaxC--\EtwoTaxD$\times$ timing advantage over the
page cache in the \texttt{pread}-based replay.  Because device traffic
is nearly equal in both configurations, that cost lies in the
page-cache lookup and reclaim path, not in which bytes each policy
keeps.  Across the
evaluated capacity and workload regimes, the kernel tier remains close
to the oracle baseline while requiring no workload-specific frequency
table, adapting to domain changes, and sharing pages across processes.
Router information is still useful for admission, HBM promotion, and
nonblocking predictive advice, while general eviction stays with the
kernel.  Two system effects require explicit control: MGLRU can
amplify traffic when most memory is unevictable, and balloon-based
capacity experiments on the tested configuration can overstate
low-capacity device traffic by approximately 2$\times$.  Reliable
evaluation therefore requires enforced capacity, per-run accounting
checks, and block-layer measurement of device traffic.

\newpage
\bibliographystyle{ACM-Reference-Format}
\bibliography{references}

\end{document}